\documentclass[a4paper,12pt]{article}
\usepackage{epsfig}
\usepackage[utf8]{inputenc}
\usepackage{amsmath,amssymb}
\usepackage{graphicx}
\usepackage{mathtext}
\usepackage{gensymb}
\usepackage[sort, numbers]{natbib}

\newcommand{\be}{\begin{equation}}
\newcommand{\ee}{\end{equation}}
\newcommand{\bea}{\begin{eqnarray}}
\newcommand{\eea}{\end{eqnarray}}
\begin{document}

\title{Strong shock in a uniform expanding universe. Approximate and exact solutions of self-similar equations}

\author{G.S. Bisnovatyi-Kogan$^{1, 2, 3}$\thanks{Email: gkogan@iki.rssi.ru},
S.A. Panafidina$^{1,3}$\thanks{Email: sofya.panafidina@phystech.edu}\\
{\small $^{1}$Space Research Institute RAS, Moscow, Russia};\\
{\small $^{2}$National Research Nuclear University MEPhI, Moscow, Russia};\\
{\small $^{3}$Moscow Institute of Physics and Technology MIPT, Dolgoprudny, Moscow reg., Russia}
}

\date{}
\maketitle

\begin{abstract}
\noindent Self-similar solution is obtained for propagation of a strong shock, in a flat expanding dusty Friedman universe. Approximate analytic solution was obtained earlier, using relation between self-similar variables, equivalent to the exact energy conservation integral, which was obtained by L.I. Sedov for the strong explosion in the static uniform medium. Numerical integration of self-similar equation is made here, giving an exact solution of the problem, which is rather close to the approximate analytic one. The differences between these solutions are most apparent in the vicinity of the shock. For polytropic equation of state, self-similar solutions exist in more narrow interval of the adiabatic power than in the static case.
\end{abstract}


\section{Introduction}

At early stages of star and galaxy formation we may expect strong explosions at last stages of evolution
of very massive primordial stars, which enrich the matter with heavy elements. Detection of heavy elements at
red shifts up to $z\sim 10$ from GRB observations (GRB090423 at $z\approx  8.2$,  GRB120923A at  $z\approx 8.5$,
GRB090429B with a photo-$z \approx 9.4$) \cite{lzgrb} make plausible this suggestion.
Propagation of a strong shock in the static uniform media was investigated by many authors \cite{stanyuk},
\cite{taylor}, but the finite analytic self-similar solution was obtained in \cite{sedov}.
Analytic self-similar solution for the strong shock propagating through the uniform expanding
 media was obtained in \cite{bk15}, describing approximately a strong shock propagation in the flat Friedman universe \cite{znuniv}. The analytic solution was obtained in neglecting the energy input from the kinetic and gravitational energy of the non-perturbed Friedman model, into heating of the matter behind the shock.
 Here the numerical solution of the same self-similar equation is obtained, which take into account all processes. It is shown that restrictions for the adiabatic power $\gamma$ obtained for validity of the analytic solution, with small corrections, remain also for the exact numerical one, and numerical difference between both solutions is not large. A qualitative behaviour of the density dependence between two solutions happens in the thin layer near the shock front.
    The problem of a strong shock propagation in the expanding medium was considered earlier in \cite{soy75,och78,be83,ito83,vob85,ks93,ek98}
Propagation of a detonation wave in the flat expanding universe was studied in \cite{kazh86,be85}. Shock propagation in the outflowing stellar wind was considered in \cite{cd89}.
Review of papers on this topic is given in \cite{omke88}.

\section{Self-similar equations for a strong shock in a uniform expanding medium }

Let us consider equations in spherical coordinates, describing in the Newtonian approximation, a uniformly expanding $v=H(t)r$, self-gravitating medium, with a density $\rho(t)$ depending only on time, what corresponds
 to the Friedman model of the universe \cite{znuniv}.

 \be
 \frac{\partial v}{\partial t}+v \frac{\partial v}{\partial r}=-\frac{1}{\rho} \frac{\partial p}{\partial r}-\frac{G_g m}{r^2},
 \quad \frac{\partial \rho}{\partial t}+ \frac{\partial \rho v}{\partial r}+\frac{2\rho v}{r}=0,
\label{eq1a}
\ee
$$
\quad
\left(\frac{\partial }{\partial t}+v \frac{\partial }{\partial r}\right)\ln{\frac{p}{\rho^\gamma}}=0,
\quad \frac{\partial m}{\partial r}={4\pi}\rho r^2.
$$
We take a flat dusty model with a velocity tending to
zero at time infinity, having density $\rho_1(t)$, and expansion velocity $v_1=H_1(t)r$. The exact solution of the system (\ref{eq1a}) for a uniform expanding self-gravitating medium is written as
\bea
\rho_1=\delta/t^2,\quad \delta=\frac{1}{6\pi G_g}, \quad \rho_1=\frac{1}{6\pi G_g t^2}; \qquad H_1=\frac{2}{3t}, \quad v_1=2r/3t;\nonumber \\
  m=\frac{4\pi}{3}\rho r^3=\frac{2r^3}{9 G_g t^2},\quad  \frac{G_g m}{r^2}=\frac{2}{9}\frac{r}{t^2}.\qquad\qquad\qquad.
\label{eq20a}
\eea
Here  $G_g$ is the gravitational constant.
The newtonian solution for the flat expan\-d\-ing universe is valid physically in the region where
$v_1\ll c_{\rm light}$, $c\ll c_{\rm light}$.
For the case of a point explosion with the energy $E$,
the number of parameters  is the same as in the previous static medium
($\delta, \,\,\, E$), therefore
we look for a self-similar solution for the case of a strong shock motion.
 The non-dimensional combination in the case of a uniformly
expanding medium is written as $r(\delta/Et^4)^{1/5}$. A position of the shock in the self-similar solution
should correspond to the fixed value of the self-similar coordinate, so that the distance of the shock
 to the center $R$ may be written as
 \be
 R=\beta\left(\frac{E t^4}{\delta}\right)^{1/5},
 \label{eq21a}
 \ee
 where $\beta$ is a number depending only on the adiabatic power $\gamma$. The velocity of the shock $u$ in the
 laboratory frame, for the unperturbed expanding gas is defined as
\be
 u=\frac{dR}{dt}=\frac{4R}{5t}=\frac{4\beta E^{1/5}}{5\delta^{1/5} t^{1/5}}.
\label{eq22a}
 \ee
The velocity of the shock $u$, the velocity of the matter behind the shock $v_2$, moving through the uniformly expanding
medium (\ref{eq20a}), are decreasing with time
$\sim t^{-1/5}$, and the pressure behind the shock $p_2$ is decreasing $\sim t^{-2/5}$,
what is considerably slower than in the case of the constant density
medium. It happens due to the fact, that the background density is decreasing with time, and the resistance to the shock
propagation is decreasing also.

The Euler equations of motion (\ref{eq20a}) are valid in this case, the conditions on the strong shock discontinuity
should be written, with account of the expansion, in the laboratory frame as
\be
 v_2=\frac{2}{\gamma+1}u+\frac{\gamma-1}{\gamma+1}v^{sh}_1,\,\, \rho_2=\frac{\gamma+1}{\gamma-1}\rho_1,\,\,
 \label{eq23a}
 \ee
$$ p_2=\frac{2}{\gamma+1}\rho_1 (u-v^{sh}_1)^2,\,\,
 c_2^2=\frac{2\gamma(\gamma-1)}{(\gamma+1)^2}(u-v^{sh}_1)^2.
 $$
Here $v_1^{sh}=\frac{2R}{3t}$ is the unperturbed expansion velocity on the shock level. The subscript "2" is related to the values behind the shock.
We introduce non-di\-men\-si\-o\-nal variables behind the shock as
 \be
 v=\frac{4r}{5t} V,\,\,\, \rho=\frac{\delta}{t^2} G,\,\,\, c^2=\frac{16r^2}{25 t^2} Z,\,\,\,m=\frac{4\pi}{3}\rho_1 r^3 M=\frac{4\pi}{3}\frac{r^3}{t^2}\delta M,
 \label{eq24a}
 \ee
depending on the self-similar variable $\xi$, defined as
 \be
\xi= \frac{r}{R(t)}= \frac{r}{\beta}\left(\frac{\delta}{Et^4}\right)^{1/5}.
 \label{eq25a}
 \ee
The conditions (\ref{eq23a}) on the strong shock at $r=R$, $\xi=1$, in non-dimensional variables (\ref{eq24a})
are written as
\be
V(1)=\frac{5\gamma+7}{6(\gamma+1)}
, \,\,\, G(1)=\frac{\gamma+1}{\gamma-1},\,\,\,
Z(1)=\frac{\gamma (\gamma-1)}{18(\gamma+1)^2},\,\,\,M(1)=1.
\label{eq26a}
\ee
In non-dimensional variables (\ref{eq24a}) the original system (\ref{eq20a}) is written as
\be
Z\left(\frac{d\ln Z}{d\ln\xi}+\frac{d\ln G}{d\ln\xi}+2\right)+\gamma(V-1)\frac{d V}{d\ln\xi}
=\gamma V(\frac{5}{4}-V)-\frac{25}{72}\gamma M,
\label{eq27a}
\ee
\be
\frac{d V}{d\ln\xi}-(1-V)\frac{d\ln G}{d\ln\xi}=-3V+\frac{5}{2},
\label{eq28a}
\ee
\be
\frac{d\ln Z}{d\ln\xi}-(\gamma-1)\frac{d\ln G}{d\ln\xi}=-\frac{5-2V-\frac{5}{2}\gamma}{1-V},
\label{eq29a}
\ee
\be
\xi\,\frac{d M}{d\xi}=3(G-M).
\label{eq29b}
\ee
Here we used relations
\be
\frac{\partial\xi}{\partial t}\bigg|_r=-\frac{4\xi}{5t},\quad  \frac{\partial\xi}{\partial r}\bigg|_t=\frac{\xi}{r}.
\label{eq30a}
\ee

\section{Construction of the approximate first integral of the problem}

 The system of four equations (\ref{eq27a})-(\ref{eq29b}), describing the self-similar solution of the problem,
 has an approximate first integral, corresponding to the energy conserva\-t\-i\-on integral in the case of the static background medium \cite{sedov}.
 In the static case without gravity the energy of the unperturbed cold medium was zero, therefore the only energy of the matter came from the energy brought by the explosion. All this was situated behind the shock, as a sum of the kinetic and thermal energy, remaining constant during the whole process. In the case of the expanding gravitating medium, it the unperturbed state it has the kinetic and gravitating energy. During the expansion after explosion they could be partially transformed into the thermal and kinetic energy of the gas behind the shock. Therefore this sum is not conserved anymore, contrary to the static case. We expect that in the case of a strong explosion the change of the explosion energy is smaller than its initial value, and may be approximately considered as a constant. This approximate integral was found in \cite{bk15}

 To find this integral, a situation was considered in the coordinate system, where the background
 medium is locally static on the shock at $r=R(t)$. In this coordinate system it is possible to
 construct the combination of functions, representing the first approximate energy integral.
  Due to self-similarity there should be conservation of this integral inside any sphere
of smaller radius $r\leq R(t)$. The  frame  which is locally comoving at $r=R$, has a velocity $v_1^{sh}=\frac{2R}{3t}$
relative to the laboratory frame, and behind the shock, at $r\leq R(t)$ we should use instead, due to self-similarity, the velocity
$v_1^{sh}\frac{r}{R}=\frac{2r}{3t}$, like in the case of the static background.
This velocity should be subtracted from all velocities in the expression for the first integral from\cite{sedov}. At $r\leq R(t)$ the velocity  $v_n=u\frac{r}{R}=\frac{4r}{5t}$.   We have than
\be
(v-v_1^{sh}\frac{r}{R})\left(\frac{c^2}{\gamma-1}+\frac{(v-v_1^{sh}\frac{r}{R})^2}{2}\right)=(v_n-v_1^{sh}\frac{r}{R})
(\frac{c^2}{\gamma(\gamma-1)}+\frac{(v-v_1^{sh}\frac{r}{R})^2}{2}).
\label{eq31a}
\ee
In the non-dimensional variables (\ref{eq24a}) we have from (\ref{eq31a})
\be
\left(\frac{4r}{5t}V-\frac{2r}{3t}\right)\left[\frac{Z}{\gamma-1}\frac{16r^2}{25t^2}
+\frac{1}{2}\left(\frac{4r}{5t}V-\frac{2r}{3t}\right)^2\right]
\label{eq32a}
\ee
$$=\left(\frac{4r}{5t}-\frac{2r}{3t}\right)
\left[\frac{Z}{\gamma(\gamma-1)}\frac{16r^2}{25t^2}
+\frac{1}{2}\left(\frac{4r}{5t}V-\frac{2r}{3t}\right)^2\right].
$$
$$v_n=\frac{4r}{5t}, \quad   v=\frac{4r}{5t}V
$$
This relation reduces to
\be
(V-\frac{5}{6})\left[\frac{Z}{\gamma-1}+(V-\frac{5}{6})^2\right]=
(1-\frac{5}{6})\left[\frac{Z}{\gamma(\gamma-1)}+(V-\frac{5}{6})^2\right],
\label{eq33a}
\ee
from what follows the first integral in the form
\be
Z=\frac{(\gamma-1)(1-V)(V-\frac{5}{6})^2}{2(V-\frac{5}{6}-\frac{1}{6\gamma})}.
\label{eq34a}
\ee
At the shock $r=R$, $\xi=1$, with $Z(1)$ and $V(1)$ from (\ref{eq26a}), the approximate first integral becomes an identity, what confirms its consistency with the problem.
Using the approximate first integral (\ref{eq34a}) we may consider only two differential equations, not containing gravity terms,
(\ref{eq28a}) and (\ref{eq29a}) for finding a solution of the problem, like in the classical Sedov case. In the case of the expanding universe the relation (\ref{eq34a}) may be interpreted as the profiling function of the temperature behind the shock. It could be possible to construct many other profiling functions, but this one is probably the only one, permitting to find the approximate analytical solution \cite{bk15}.

\section{Approximate analytic solution for expan\-d\-ing\\ me\-di\-um}

Similar to the static background medium in the previous section (see \cite{sedov},\cite{sedov1}),
we look for an analytic solution of (\ref{eq28a}),(\ref{eq29a}), where
the variable $Z$ is determined by the algebraic relation (\ref{eq34a}).
Excluding $\frac{d\ln G}{d\ln\xi}$ from (\ref{eq28a}),(\ref{eq29a}), we obtain the equation
\be
(\gamma-1)\frac{dV}{d\ln\xi}-(1-V)\frac{d\ln Z}{d\ln\xi}=\frac{5}{2}-3\gamma V+V.
\label{eq35a}
\ee
From (\ref{eq34a}) we find
\be
\frac{d\ln Z}{d\ln\xi}=-\frac{1}{1-V}\,\frac{dV}{d\ln\xi}+\frac{2}{V-\frac{5}{6}}\,\frac{dV}{d\ln\xi}
-\frac{\gamma}{\gamma V-\frac{5\gamma}{6}-\frac{1}{6}}\,\frac{dV}{d\ln\xi}.
\label{eq36a}
\ee
After substituting in (\ref{eq35a}) we find the following equation for the variable $V$
\be
\left(\gamma+1-\frac{2}{6V-5}+\frac{\gamma-1}{6\gamma V-5\gamma
-1}\right)\frac{d V}{d\ln \xi}=\frac{5}{2}-3\gamma V+V.
\label{eq37a}
\ee
The solution of Eq. (\ref{eq37a}) is written in the form \cite{bk15}
\be
\label{39a}
 -\frac{2}{20-15\gamma}\ln(V-\frac{5}{6})
+\frac{\gamma-1}{17\gamma-15\gamma^2+1}\ln(V-\frac{5}{6}-\frac{1}{6\gamma})
\ee
$$
+\left[-\frac{\gamma+1}{3\gamma-1}-\frac{\gamma-1}{17\gamma-15\gamma^2+1}
+\frac{2}{20-15\gamma}\right]\ln\left(V-\frac{5}{6\gamma-2}\right)
=\ln\xi+{\rm const}.
$$
Using the boundary condition (\ref{eq26a}) for $\xi=1$, we obtain the solution for $V(\xi)$ as
\be
\label{eq40a}
\left[(\gamma+1)(3V-\frac{5}{2})\right]^{\mu_1}
\left[\frac{\gamma+1}{\gamma-1}(6\gamma V-5\gamma-1)\right]^{\mu_2}
\left[6(\gamma+1)\frac{3\gamma V-V-\frac{5}{2}}{15\gamma^2+\gamma-22}\right]^{\mu_3}=\xi,
\ee
with
\be
\mu_1= \frac{2}{15\gamma-20},\,\,\, \mu_2=\frac{\gamma-1}{17\gamma-15\gamma^2+1},
\label{eq41a}
\ee
$$\mu_3=-\frac{\gamma+1}{3\gamma-1}-\frac{\gamma-1}{17\gamma-15\gamma^2+1}
+\frac{2}{20-15\gamma}.$$

For finding a solution for $G(\xi)$ we write the equations (\ref{eq28a}) and (\ref{eq37a}) in the form
\be
1-(1-V)\frac{d \ln G}{d V}=-(3V-\frac{5}{2}) \frac{d \ln\xi}{d V},
\label{eq42a}
\ee
\be
\frac{d \ln\xi}{d V}=\frac{\gamma+1-\frac{2}{6V-5}+\frac{\gamma-1}{6\gamma V-5\gamma-1)}}{\frac{5}{2}-V(3\gamma-1)}.
\label{eq43a}
\ee
The equation for $G(V)$ is written in the form
\be
1-(1-V)\frac{d \ln G}{d V}=-(3V-\frac{5}{2})\frac{\gamma+1-\frac{2}{6V-5}+
\frac{\gamma-1}{6\gamma V-5\gamma-1}}{\frac{5}{2}-V(3\gamma-1)}.
\label{eq44a}
\ee
The solution of (\ref{eq44a}) has a form
\be
\ln G=\kappa_1\ln(1-V))+\kappa_2\ln(V-\frac{5\gamma+1}{6\gamma})
\label{eq49a}
+\kappa_3\ln(V-\frac{5}{6\gamma-2})+{\rm\bf const_1}.
\ee
Here
$$
\kappa_1=\frac{7}{3\gamma-1}-\frac{2}{6\gamma-7}+\frac{(15\gamma-20)(\gamma-1)}{(6\gamma-7)(15\gamma^2-17\gamma-1)}
$$
\be
\label{eq50a}
-\frac{3\gamma(15\gamma-20)}{(3\gamma-1)(15\gamma^2-17\gamma-1)}-\frac{15\gamma-20}{3\gamma-1}\,\frac{\gamma+1}{6\gamma-7},
\ee
$$
\kappa_2=-\frac{3}{3\gamma-1}+\frac{3\gamma(15\gamma-20)}{(3\gamma-1)(15\gamma^2-17\gamma-1)}.
$$
$$
\kappa_3=\frac{2}{6\gamma-7}-\frac{(15\gamma-20)(\gamma-1)}{(6\gamma-7)(15\gamma^2-17\gamma-1)}
+\frac{15\gamma-20}{3\gamma-1}\,\frac{\gamma+1}{6\gamma-7},
$$
The {\bf const$_1$} is found from the boundary conditions (\ref{eq26a}), and finally we obtain the solution for $G(V)$ in the form
\be
\label{eq51a}
G(V)=\frac{\gamma+1}{\gamma-1}\left[6\frac{(\gamma+1)(1-V)}{\gamma-1}\right]^{\kappa_1}
\left[\frac{\gamma+1}{\gamma-1}(6\gamma V-5\gamma-1)\right]^{\kappa_2}
\ee
$$
\times
\left(\frac{3(\gamma+1)}{15\gamma^2+\gamma-22}[(6\gamma-2)V-5)]\right)^{\kappa_3}.
$$
The function $Z(V)$ is determined by the integral (\ref{eq34a}).

The mass function $M(\xi)$ is obtained from the solution of Eq. (\ref{eq29b}), which has a form of a linear non-uniform equation. The uniform equation $\xi\,\frac{d M}{d\xi}=-3M$ has a general solution $M=C\,\xi^{-3}$. Looking for the particular solution of the non-uniform equation in the form $M=C(\xi)\,\xi^{-3}$, we obtain $C(\xi)=3\int_0^\xi G(\eta)\eta^2 d\eta$. Finally. the nonsingular solution for $M$ is written as
\be
M(\xi)=3\,\xi^{-3}\,\int_0^\xi G(\eta)\eta^2 d\eta.
\label{eq52b}
\ee

\section{Main properties of the approximate analy\-tic so\-lu\-ti\-on}

The analytic solution  (\ref{eq40a}),(\ref{eq51a}),(\ref{eq34a}),(\ref{eq52b}) has a complicated dependence of $\gamma$, and physically relevant solution exist only for limited values of $\gamma$. To have positive values in brackets of (\ref{eq40a}), and to satisfy the condition  for $V$ on the shock (\ref{eq26a}) we obtain restrictions for $V$ as
\be
V>\frac{5}{6},\quad V>\frac{1+5\gamma}{6\gamma}, \quad V<V(1)=\frac{5\gamma+7}{6(\gamma+1)}.
\ee
To satisfy all these conditions we obtain the restriction for $\gamma$ as $1<\gamma<\gamma_*$, where
$\gamma_*$ is defined by equation
\be
\label{eq53b}
15\gamma^2+\gamma-22=0, \qquad \gamma_*=-\frac{1}{30}+\sqrt{\frac{1}{900}+\frac{22}{15}},\qquad
 \gamma_*\approx 1.1782.
\ee
Numerical solution of self-similar equations (\ref{eq27a})-(\ref{eq29b}), presented below, has very similar restrictions for $\gamma$. We may conclude, therefore, that for other $\gamma>\sim\gamma_*$ there are no self-similar solutions. On figures are plotted, for different $\gamma<\gamma_*$,  functions from the analytical solution: $V(\xi)$  from (\ref{eq40a}) in Fig.1; $G(\xi)$ from (\ref{eq51a}) in Fig.2; $Z(\xi)$ from (\ref{eq34a}) in Fig.3; and $M(\xi)$ from (\ref{eq52b}) in Fig.4.

\begin{figure}
\center{\includegraphics[width=1\linewidth]{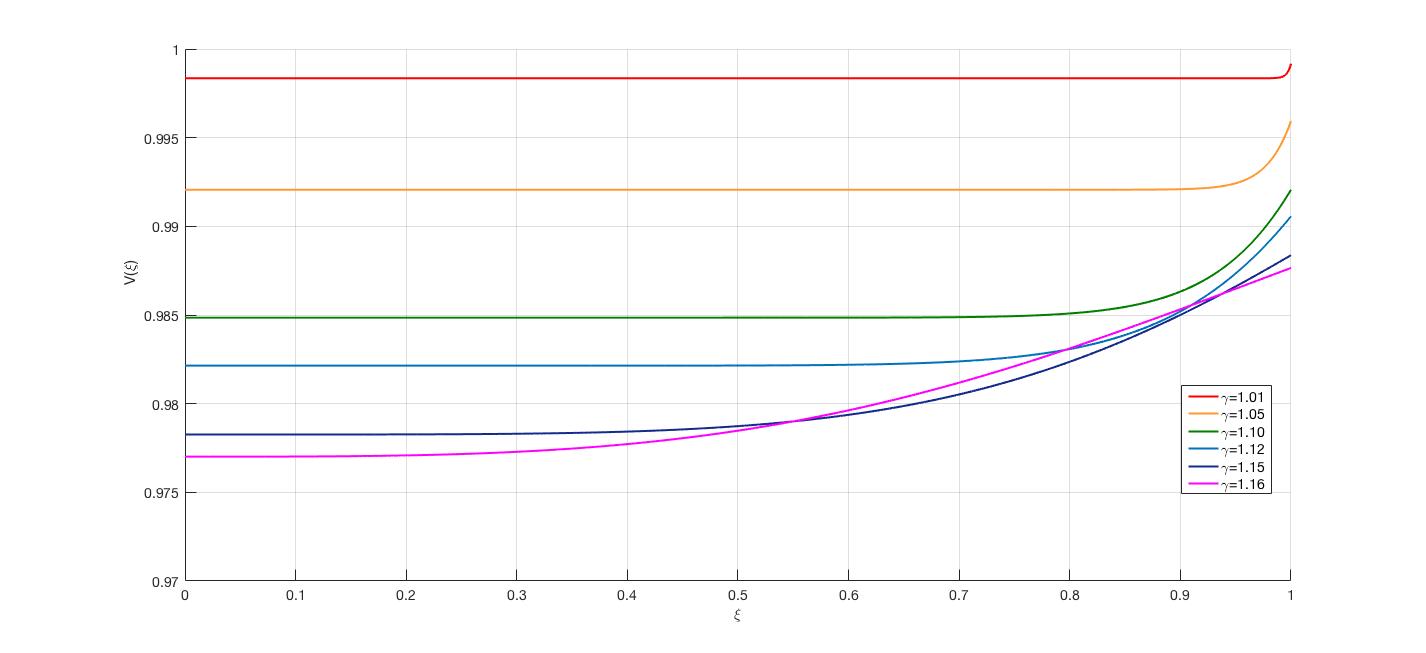}}
\caption{Approximate analytic solution for $V(\xi)$}
\label{ris:image}
\end{figure}

\begin{figure}
\center{\includegraphics[width=1\linewidth]{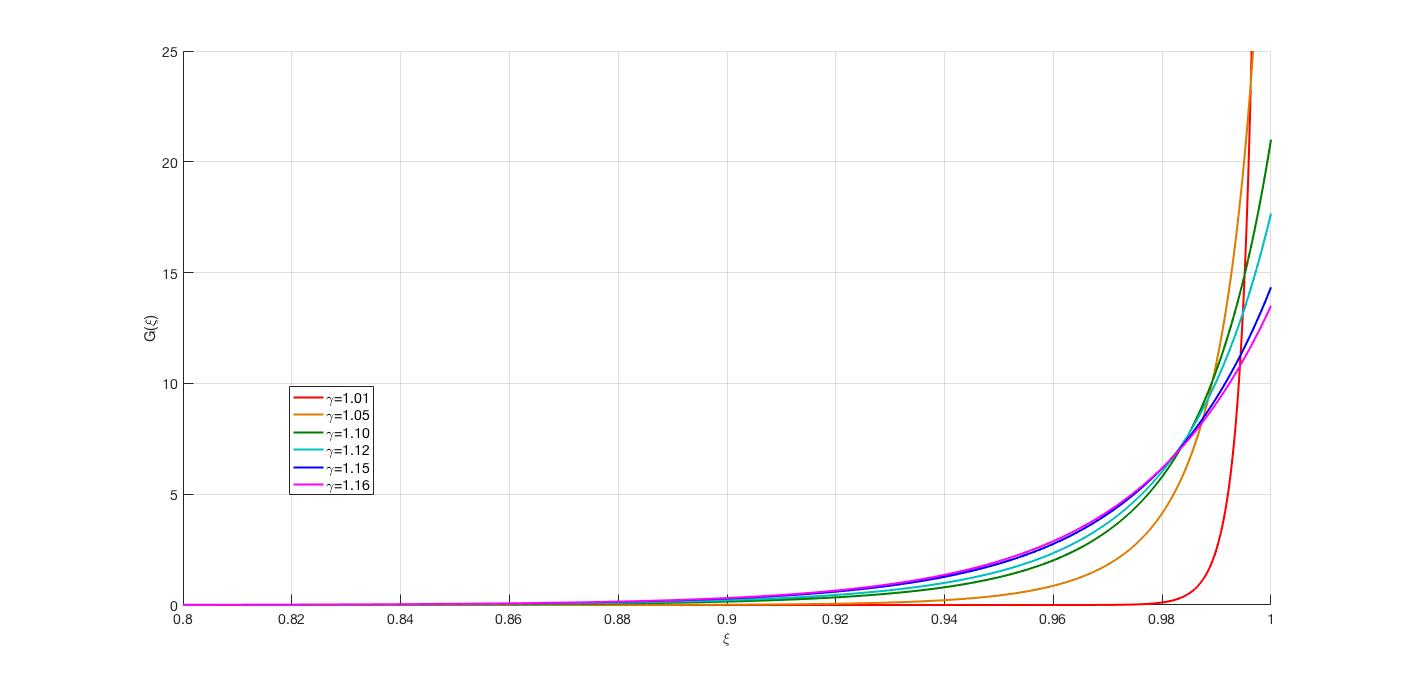}}
\caption{Approximate analytic solution for $G(\xi)$}
\label{ris:image}
\end{figure}

\begin{figure}
\center{\includegraphics[width=1\linewidth]{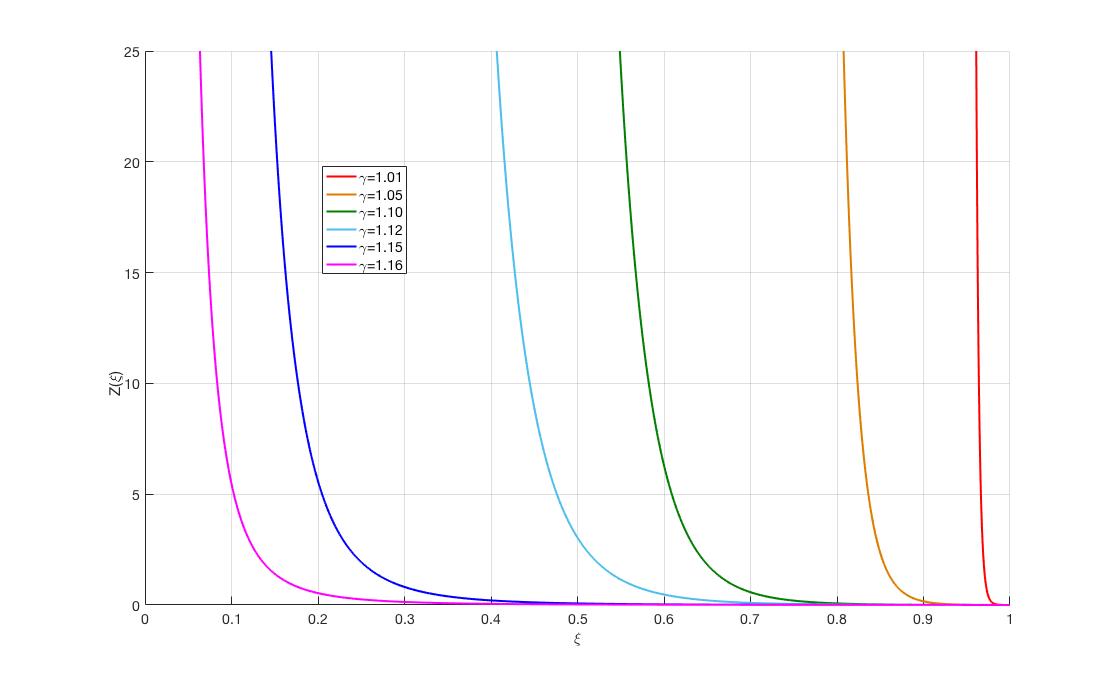}}
\caption{Approximate analytic solution for $Z(\xi)$}
\label{ris:image}
\end{figure}

\begin{figure}
\center{\includegraphics[width=1\linewidth]{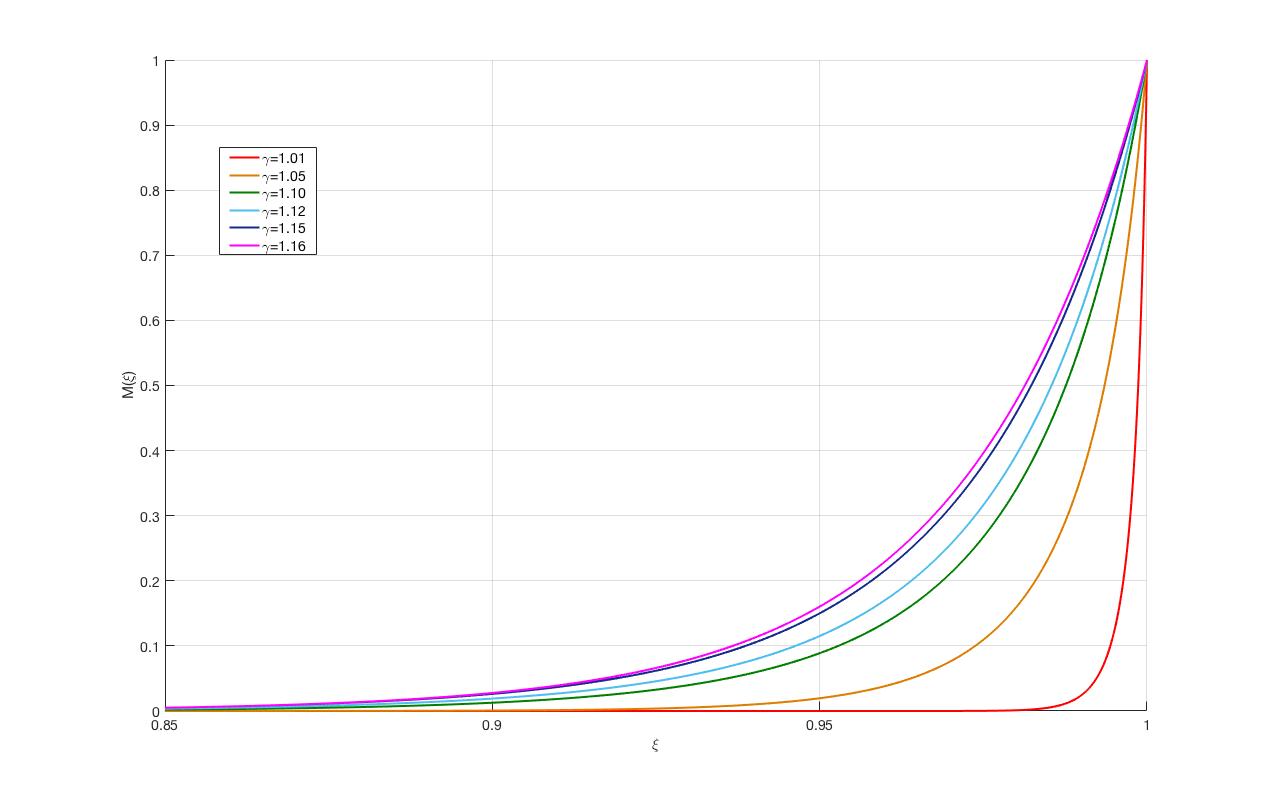}}
\caption{Solution for $M(\xi)$ from (\ref{eq52b}) based on approximate analytic equations}
\label{ris:image}
\end{figure}

Introduce notations
\be
\label{eq54b}
 V'=\frac{d\,V}{d\,\xi},\quad 
 G'=\frac{d\,G}{d\,\xi},\quad
 Z'=\frac{d\,Z}{d\,\xi}
\ee
At the shock $\xi=1$ the derivative of the self-similar functions are found from (\ref{eq36a}),(\ref{eq37a}),(\ref{eq28a}) giving
\be
 V'(1)=\frac{-15\gamma^2-\gamma+22}{6(\gamma+1)^2};\quad
 G'(1)= \frac{-15\gamma^2+5\gamma+28}{(\gamma-1)^2};
\quad Z'(1)= \frac{(15\gamma^2+\gamma-22)\gamma}{9(\gamma+1)^3}
\label{eq54c}
\ee
It follows from (\ref{eq53b}),(\ref{eq54c}), that for $\gamma<\gamma_*$ the derivative have the following signs
\be
 V'(1)>0;\quad G'(1)>0;\quad  Z'(1)<0
\label{eq54d}
\ee

\section{Numerical solution of self-similar equations}

The system of equations (\ref{eq27a})-(\ref{eq29b}), written explicitly for deriva\-ti\-v\-es, has a form:  \\
\begin{equation*}
\label{eq54c}
\begin{cases}
$$\frac{dlnG}{dln\xi}=\frac{\frac{3-\frac{5}{2}\gamma}{1-V}Z-\frac{25}{72}\gamma M   +\gamma(2V^2-\frac{17}{4}V+\frac{5}{2})}{\gamma[Z-(1-V)^2]};$$\\
$$\frac{dV}{dln\xi}=(1-V)\frac{dlnG}{dln\xi}-3V+\frac{5}{2};$$\\
$$\frac{dlnZ}{dln\xi}=(\gamma-1)\frac{dlnG}{dln\xi}-\frac{5-2V-\frac{5}{2}\gamma}{1-V};$$\\
$$\frac{dM}{dln\xi}=3(G-M)$$ \\
\end{cases}
\end{equation*}
That reduces to:
\be
\label{eq55b}
\xi\frac{dG}{d\xi}=G\frac{\frac{3Z}{\gamma}\frac{1-\frac{5\gamma}{6}}{1-V}-\frac{17}{4}V+\frac{5}{2}+2\,V^2
-\frac{25}{72}M}{Z
-(1-V)^2},\quad \xi\,\frac{d M}{d\xi}=3(G-M),
\ee
\be
\nonumber
\xi\frac{dV}{d\xi}=\xi\frac{1-V}{G}\frac{dG}{d\xi}-3(V-\frac{5}{6}),\quad
\frac{\xi}{Z}\frac{dZ}{d\xi}=\xi\frac{\gamma-1}{G}\frac{dG}{d\xi}-\frac{5-2V-\frac{5}{2}\gamma}{1-V}.
\ee

Note, that the expression (\ref{eq52b}) for $M(\xi)$ is valid also for the exact numerical solution.
 This system is solved by integration, starting from the point $\xi=1$, where the variables are defined by conditions on the shock (\ref{eq26a}). At this boundary the derivatives have the following values

\be
\label{eq57b}
\quad \frac{dV}{d\xi}\bigg|_{\xi=1}=\frac{-30\gamma^2-11\gamma+27}{6(\gamma+1)^2};\quad \frac{dG}{d\xi}\bigg|_{\xi=1}=\frac{-30\gamma^2-5\gamma+33}{(\gamma-1)^2};
\ee
\be
\nonumber
\frac{dZ}{d\xi}\bigg|_{\xi=1}= -\frac{\gamma(15\gamma^3-35\gamma^2-17\gamma+49)}{18(\gamma+1)^3};\quad 
\frac{dM}{d\xi}\bigg|_{\xi=1}=\frac{6}{\gamma-1}
\ee
Here derivatives $V'$, $G'$ and $Z'$ are negative at $\xi=1$, what differs from the derivatives in the approximate analytic solution in (\ref{eq54d}). Numerical integration of the system (\ref{eq55b}) shows, that close to the shock boundary the values of $G(\xi)$ and $V(\xi)$ reach their maxima, and decrease monotonically  until the origin $\xi=0$, see Figs. 5,6,7. Numerical solutions for $Z(\xi)$ and $M(\xi)$ for different $\gamma$ are given in Figs.8,9, respectively. The self-similar solutions, filling all space, exist only in the interval 
$1<\gamma<\gamma_{**}$, where $\gamma_{**}=1.155$.
 For $\gamma > \gamma_{**}=1.155$ the empty spherical space should be formed 
around the center, at a finite distance behind the shock,
similar to  Sedov solution for the shock in the static uniform gas, where the empty 
sphere around the center is formed 
at $\gamma > 7$ \cite{llhydro}.

\begin{figure}
\center{\includegraphics[width=1\linewidth]{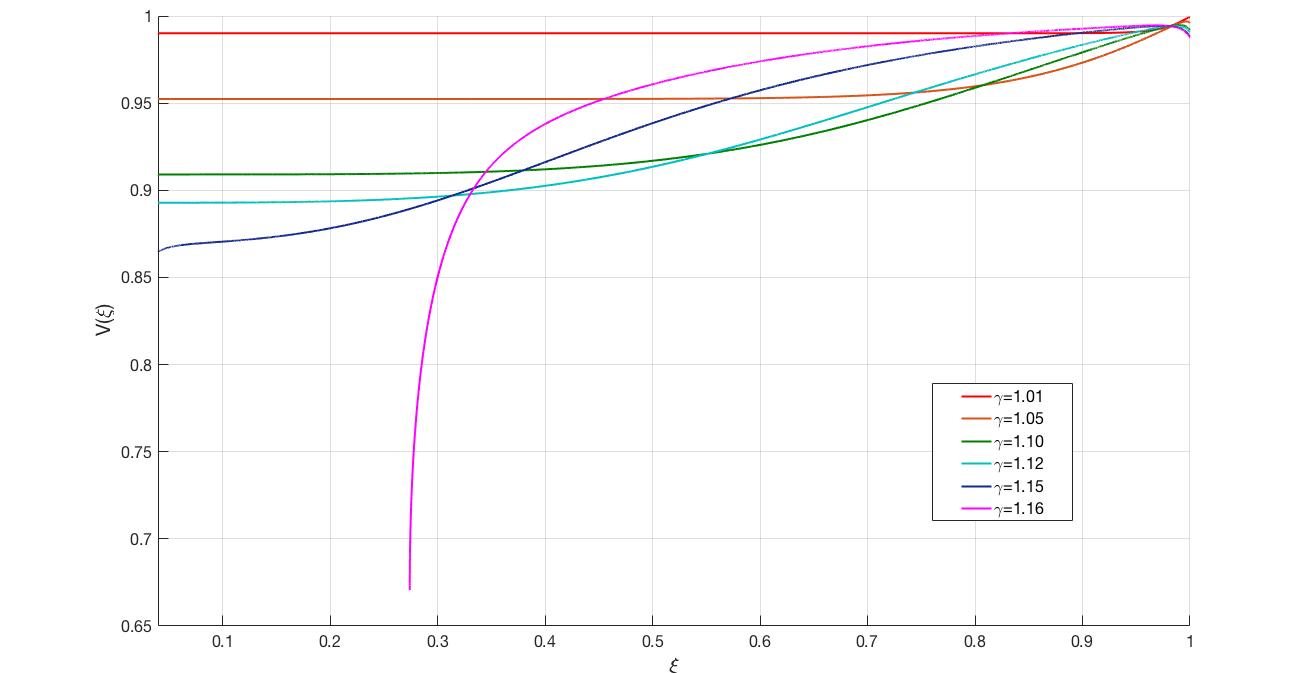}}
\caption{Numerical solution for $V(\xi)$}
\label{ris:image}
\end{figure}

\begin{figure}
\center{\includegraphics[width=1\linewidth]{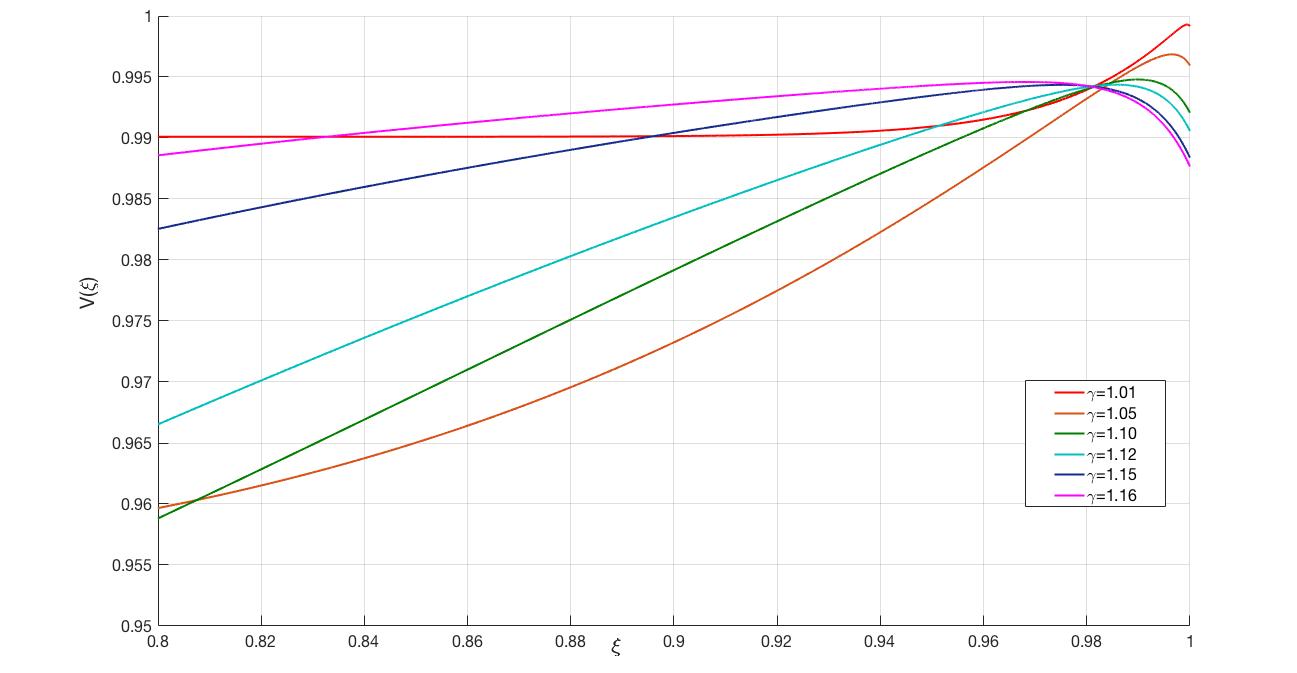}}
\caption{Numerical solution for $V(\xi)$ where $\xi$ from 0.8 to 1.0}
\label{ris:image}
\end{figure}

\begin{figure}
\center{\includegraphics[width=1\linewidth]{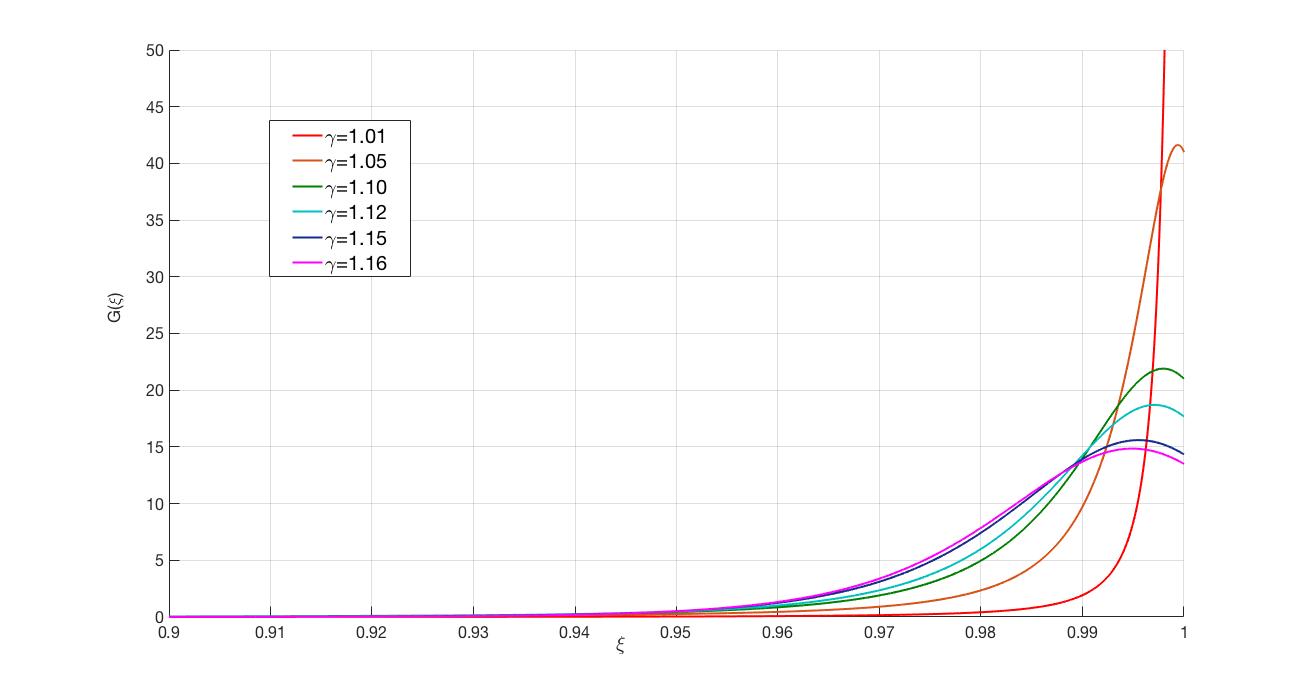}}
\caption{Numerical solution for $G(\xi)$ where $\xi$ from 0.9 to 1.0}
\label{ris:image}
\end{figure}

\begin{figure}
\center{\includegraphics[width=1\linewidth]{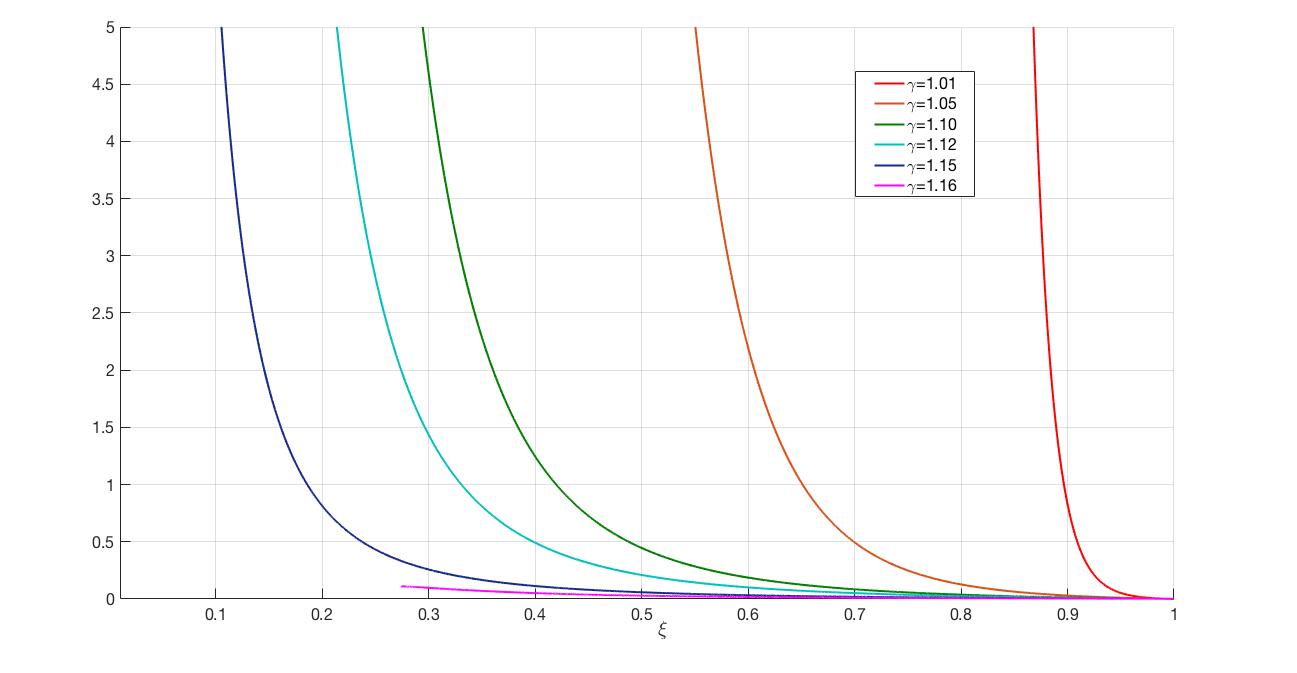}}
\caption{Numerical solution for $Z(\xi)$}
\label{ris:image}
\end{figure}

\begin{figure}
\center{\includegraphics[width=1\linewidth]{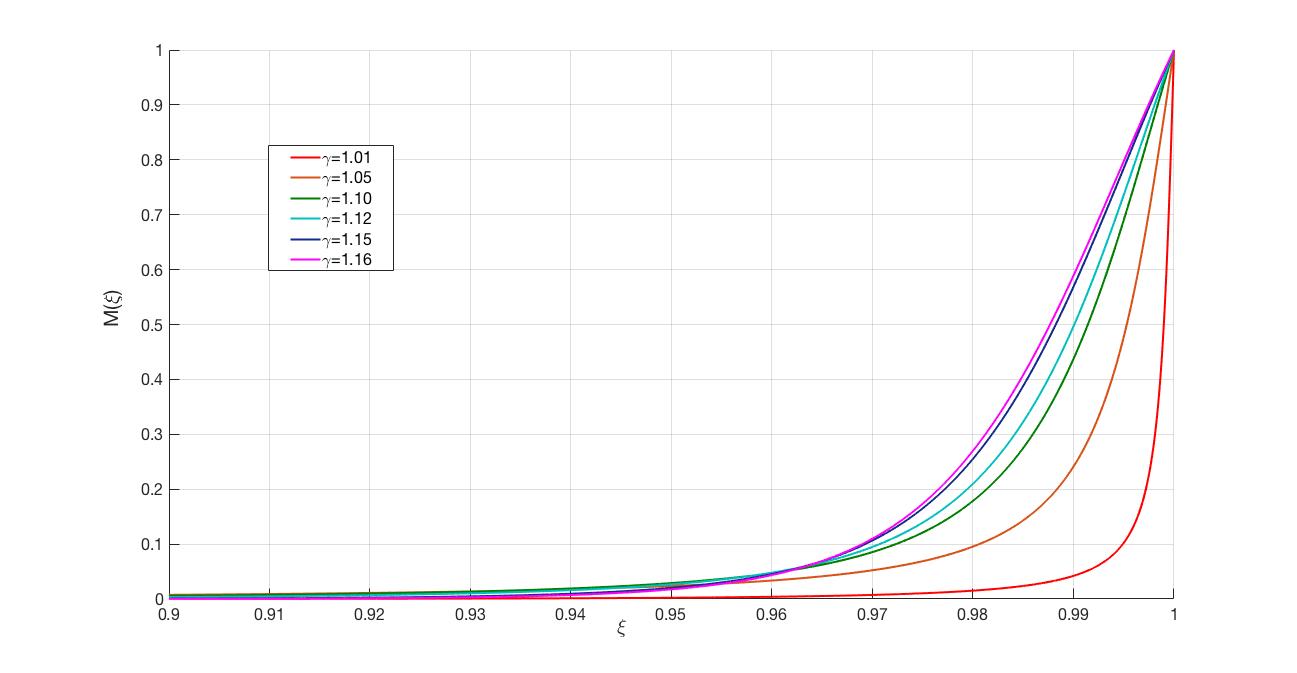}}
\caption{Numerical solution for $M(\xi)$ where $\xi$ from 0.9 to 1.0}
\label{ris:image}
\end{figure}

\section{Discussion}

The constant $\beta$ in the definition of the non-dimensional radius $\xi$ in (\ref{eq25a}) is obtained from the
explosion energy integral $E$. Due to zero energy (kinetic + gravitational) in the non-perturbed solution, the conserving value of the explosion energy behind the shock, in
the uniformly expanding medium, with velocity and density distribu\-ti\-ons (\ref{eq20a}), with account of the gravitational energy, is determined as
\be
\label{eq52a}
E=\int_0^{R(t)} \rho\left[\frac{v^2}{2}+\frac{c^2}{\gamma(\gamma-1)}\right]4\pi r^2 dr-
\int_0^{R(t)}\frac{G_g m dm}{r}.
\ee
In non-dimensional variables (\ref{eq24a}) this relation reduces to the equation for the constant $\beta$
\be
\label{eq53a}
\beta^{-5}=\frac{64\pi}{25}\int_0^1 G\left[\frac{V^2}{2}+\frac{Z}{\gamma(\gamma-1)}\right]\xi^4 d\xi-\frac{8}{3}\int_0^1 G\xi\left(\int_0^\xi G\eta^2 d\eta\right)d\xi.
\ee

\begin{table}[h]
\caption{The values $\beta(\gamma)$ for the analytic and numerical solutions}
\label{tabular:timesandtenses}
\begin{center}
\begin{tabular}{ |c |c |c|}
\hline
\textbf{$\gamma$} & \textbf{$\beta_{an}$} & \textbf{$\beta_{num}$} \\
\hline
1.05 &  3.2910 & 3.3512 \\
\hline
1.10 & 2.2268 & 2.5003 \\
\hline
1.12 & 2.0423 & 2.3713 \\
\hline
1.15 & 1.8522 & 2.2416 \\
\hline
\end{tabular}
\end{center}
\end{table}

The values $\beta(\gamma)$ for the analytic solution at $\gamma<\gamma_{**}$, and numerical one for $\gamma<\gamma_{**}$, are given in the Table.
It follows from the self-similar solution, that in the expanding medium the velocity of shock from (\ref{eq22a})
decreases as $\sim t^{-1/5}$, what is much slower than the shock veloci\-ty in the static uniform medium $\sim t^{-3/5}$, according
to Sedov solution \cite{sedov}. Correspondingly the radius of the shock wave in the expanding self-gravita\-t\-ing medium
increases $\sim t^{4/5}$, more rapidly that the shock wave radius in the uniform non-gravitating medium $\sim t^{2/5}$.
It means, that the shock propagates in the direction of decreasing density with larger speed, than in the static medium,
 due to accele\-ra\-t\-ing action
of the decreasing density, even in the presence of a self-gravitation.

\section*{Acknowledgments}

  This work  was partially supported by
  RFBR grants 17-02-00760, 18-02-00619, and RAN Program No.28 "Astrophysical objects as cosmic laboratories".
    Authors are grateful to I. Kovalenko for important bibliographical references.

\end{document}